%% file: main.tex
\documentclass{stsci_report}

\usepackage{graphicx}
\usepackage{listings}
\usepackage{xcolor}

\usepackage{siunitx}
\usepackage{todonotes}
\usepackage{pdfpages}
\usepackage{hyperref}
\usepackage{dirtytalk}
\usepackage[section]{placeins}
\usepackage{caption}
\usepackage{aas_macros}

\usepackage{float}
\usepackage{graphicx}
\usepackage{siunitx}
\usepackage{todonotes}
\usepackage{pdfpages}
\usepackage{tabularx}
\usepackage{multirow}
\usepackage[indent]{parskip}
\usepackage{subcaption}
\usepackage{aas_macros}
\usepackage[bottom]{footmisc}
\usepackage{wrapfig}
\usepackage{amsmath}
\usepackage{colortbl}
\usepackage{tabularray }
\usepackage{setspace} 
\usepackage{natbib}
\usepackage{listings}
\usepackage{ragged2e}

\DeclareCaptionType{equ}[][]
\usepackage{amsmath} 

\definecolor{codegreen}{rgb}{0,0.6,0}
\definecolor{codegray}{rgb}{0.5,0.5,0.5}
\definecolor{codepurple}{rgb}{0.58,0,0.82}
\definecolor{backcolour}{rgb}{0.95,0.95,0.92}

\lstdefinestyle{mystyle}{
  backgroundcolor=\color{backcolour},   
  commentstyle=\color{blue},
  keywordstyle=\color{codegreen},
  numberstyle=\tiny\color{codegray},
  stringstyle=\color{codepurple},
  basicstyle=\ttfamily\footnotesize,
  breakatwhitespace=false,         
  breaklines=true,                 
  captionpos=b,                    
  keepspaces=true,                 
  numbers=left,                    
  numbersep=5pt,                  
  showspaces=false,                
  showstringspaces=false,
  showtabs=false,                  
  tabsize=1
}

\copyrighttext{Copyright\copyright\ \the\year\ The Association of Universities for Research in Astronomy, Inc. All Rights Reserved.}

\presubtitle{ISR 2024-15}
\title{WFC3/UVIS Geometric Distortion - Time Evolution of Linear Terms w.r.t Gaia}
\author{Anne O'Connor, Varun Bajaj, Jennifer Mack, Annalisa Calamida}
\date{December 30, 2024}

\begin{document}

\maketitle
\begin{Center}
\abstract{We align more than 7,400 WFC3/UVIS exposures to the Gaia DR3 catalog to examine the time evolution of the linear terms (shift, rotation, scale and skew) of the geometric distortion solution between 2009 and 2022. We find small linear temporal changes in the scale and skew terms (less than $0.2$ pixels in 13 years) which are generally dominated by intrinsic scatter (up to $\pm 0.3$ pixels). Concurrently, a larger filter-dependent offset in the scale term is observed, with a maximum difference of 0.3 pixels between F275W and F814W images at all epochs. A small rotation offset to Gaia of $0.003 \pm 0.004$ degrees is measured from 2009 to mid-2017, after which the offsets are as large as 0.01 degrees, with a large scatter. MAST pipeline processing includes an additional alignment step which corrects UVIS images for any residual linear terms with respect to Gaia DR3 when there are at least 10 matched sources. In addition to any pointing offsets, this step accounts for any evolution in the distortion linear terms described here. For observers requiring high-precision astrometry, we recommend using the \texttt{tweakreg}\footnote{Part of the \href{https://drizzlepac.readthedocs.io/en/latest/}{DrizzlePac} package. } routine to realign images using a 4-parameter fit ($x-$shift, $y-$shift, rotation, and scale) or a 6-parameter fit ($x-$shift, $y-$shift, $x-$rotation, $y-$rotation, $x-$scale, and $y-$scale) depending on the number of matched sources. We provide links to DrizzlePac tutorials for improving both absolute and relative astrometry in WFC3 images.
}
\end{Center}

\section{Introduction}
\label{sec:Intro}
\subsection{WFC3/UVIS Geometric Distortion}
\label{sec:geo_dist intro}
It is important to accurately characterize the geometric distortion of the WFC3/UVIS detector in order to derive precise sky coordinate positions, parallaxes, and proper motions.  Additionally, an accurate geometric distortion model is essential for correcting the images to allow mosaicking and stacking (drizzling) into a common frame.

Previous studies have precisely characterized the geometric distortion model with a 4th-order polynomial in x- and y-coordinates, as described in WFC3 Instrument Science Report (ISR) 2009-33 \citep{kozhurina2009wfc3}. These solutions were delivered to the WFC3 calibration pipeline as an Instrument Distortion Coefficients Table (IDCTAB).  Additionally, residual non-polynomial distortions are stored in lookup tables (NPOLFILE and D2IMFILE) and are used by the pipeline to remove any remaining distortion.  These files are used to create the World Coordinate System (WCS) solution of a given image; the WCS information is essential in astronomical data analysis since it is needed for creating source catalogs and for combining images, i.e. with AstroDrizzle \citep{hoffmann2021drizzlepac}, DrizzlePac’s
primary user interface to create distortion-corrected, cosmic-ray cleaned, and combined images.

The distortion solutions for WFC3/UVIS filters are stored using the FITS WCS standard \citep{fits_wcs1}.  The WCS and the included distortions are used to transform the image pixel coordinates into sky coordinates. The core WCS keywords in the header of images store the linear components of the geometric transformation: shift in RA/Dec, rotation of the $y-$axis with respect to North, pixel $x-$scale, pixel $y-$scale, and skew between the x and y axes, with nonlinear components stored using the Simple Imaging Polynomial convention \citep{fits_sip}. 

To ensure that WFC3/UVIS images have precise astrometric information, it is important to evaluate the effectiveness of the geometric distortion solution over time. Significant changes to the linear terms of the geometric distortion could result in blurred and distorted sources in drizzled images when combining multi-epoch datasets with Astrodrizzle. This distortion, leading to inaccurate astrometry, could result in decreased precision of scientific measurements such as proper motions.

Past studies investigated the time-dependence of the WFC3/UVIS geometric distortion. \cite{kozhurina2012time} concluded that the WFC3/UVIS geometric distortion was stable over the initial two years of WFC3 operation, except for short term variations possibly due to orbital breathing. Similarly, \cite{kozhurina2015standard} used a newly created standard astrometric catalog of the globular cluster Omega Centauri ($\omega$ Cen) to examine the linear part of the WFC3/UVIS distortion solution, and found that the WFC3/UVIS geometric distortion was time-independent and accurate at the level of ±2 mas over the 5-year time period of WFC3 operations, with no sudden or extreme fluctuations in the WFC3/UVIS astrometric scale. Another set of studies characterized the accuracy of the HST Standard Astrometric Reference Catalog (of the globular cluster $\omega$ Cen) used to derive the current distortion solution to various Gaia releases (\citealt{2016}, \citealt{kozhurina2018accuracy}, \citealt{martlin2019comparison}, \citealt{kozhurina2021accuracy}).   However, these studies were limited by their use of just a single target, $\omega$ Cen, which was not observed at the necessary fidelity or cadence in the Gaia DR1 and DR2 catalogs.\footnote{\href{https://www.cosmos.esa.int/web/gaia/dr1}{Gaia Data Release 1}, \href{https://www.cosmos.esa.int/web/gaia/data-release-2}{Gaia Data Release 2}.} Gaia DR3 \citep{2023A&A...674A...1G} and a recent Focused Product Release\footnote{\href{https://www.cosmos.esa.int/web/gaia/fpr}{Gaia Focused Product Release}} has since increased the precision and completeness of the measurements for this target.  However, as Gaia is an all-sky catalog, many other regions of the sky observed with WFC3/UVIS can be used for astrometric calibration verification.

This report presents a novel method of evaluating the change in WFC3/UVIS linear distortion terms over time with respect to the Gaia DR3 catalog.  By aligning many images over a long baseline and recording the transformations used to best align the images, we measure errors in the combined UVIS distortion solution (IDCTAB, NPOLFILE, and D2IMFILE) and how these errors change over time. In this work, we test only the linear terms and assume the higher order polynomial and residual non-polynomial terms to be constant.

\section{Data and Methodology}
\label{sec:Data}
\subsection{UVIS Data Sets For Testing}
\label{sec:uvis_data}

In order to analyze a large sample of images, we use every full-frame WFC3/UVIS exposure collected between 2009 and 2022 in 5 commonly used wide-band filters (F275W, F336W, F438W, F606W, and F814W). We discard subarray, binned or moving target images, failed observations, and any observations containing fewer than 50 Gaia sources. We then generate source catalogs for each image and use these catalogs to align each image to the Gaia DR3 catalog. The resulting set of WFC3/UVIS images is then further trimmed to include only well-aligned images. This process is described in more detail in the following sections. The resulting dataset includes a total of 7,491 WFC3 UVIS images (see Table \ref{tab:num_images_by_filter}). 
\input{tables/Data_by_filter}

Given the large number of images of a variety of different astronomical fields in this sample, possible systematic errors that could arise with a small sample size or non-uniform placement of sources on the detector are minimized.  The sample covers nearly the entirety of WFC3's lifetime aboard Hubble, allowing for a long baseline to study the time evolution of the geometric distortion.

\subsection{Methodology}
\label{sec:methodology}

\subsubsection{Generating Source Catalogs for UVIS and Gaia}
\label{sec:catalogs}

To effectively evaluate the stability of the linear terms of the WFC3/UVIS geometric distortion solution over the instrument's lifetime, we compute the linear transformations between WFC3/UVIS and Gaia for a large, diverse set of images. To build the WFC3/UVIS data sets for testing, the individual FLC files are first run through the \texttt{hst1pass} routine \citep{anderson2022one}, which detects sources on the WFC3/UVIS images and performs point spread function (PSF) photometry on the sources. A catalog containing high-precision pixel positions and magnitudes (fluxes), as well as fit quality metrics for the detected sources is generated.

Next, the \texttt{hst1pass} catalogs for each image are run through a pipeline designed to align HST images to the Gaia DR3 catalog. In the pipeline, a Gaia query is performed by providing a coordinate outline of the \say{footprint} of the image on the sky. These coordinates are used to define a bounding box for a query provided to the \texttt{astroquery.gaia} software \citep{2019AJ....157...98G}, returning a catalog of Gaia sources within the footprint. The routine also corrects the Gaia source positions for proper motion between the Gaia epoch (2016.0) and observation date. Finally, full-frame exposures with more than 50 Gaia sources are matched and aligned to Gaia using the \texttt{align\textunderscore wcs} routine from the \texttt{tweakwcs} package \footnote{Please see the \href{https://tweakwcs.readthedocs.io/en/latest/}{\texttt{tweakwcs} Documentation} for more information.}.

\subsubsection{Aligning WFC3/UVIS to Gaia}
\label{sec:alignment}

The \texttt{tweakwcs} (and \texttt{tweakreg}) routine allows a user to input two different catalogs of sources and compute the astrometric transformations between them: in this case, the WFC3/UVIS image catalog (from \texttt{hst1pass}) and its corresponding Gaia reference catalog. The alignment process happens in two stages; first, sources are cross-matched between the catalogs, then an affine transform between matched source positions is calculated. \texttt{Tweakwcs} creates an \textbf{undistorted} reference tangent plane based on the input data, with a corresponding undistorted WCS. 

All calculations are performed in pixel space in this reference tangent plane. The matching step employs the \texttt{XYXYMatch}\footnote{
For more information, see the 
\href{https://tweakwcs.readthedocs.io/en/latest/source/matchutils.html\#tweakwcs.matchutils.XYXYMatch}
{\texttt{tweakwcs} documentation page}
} algorithm, which shifts the positions of sources in the image catalog (projected into the tangent plane) in small steps in $x$ and $y$. The $x$- and $y$- shift values that result in the largest number of sources coincident with the reference catalog (in the tangent plane) are used as an initial guess for the affine transform shifts. The image catalog sources with a corresponding reference catalog source within a specified distance threshold (after the shifts are applied) are considered matches, and their corresponding positions are recorded.

Following the matching, the astrometric transformations are computed to minimize the distance between the matched image catalog positions and their corresponding reference catalog positions in the tangent plane. The following equations are used by \texttt{tweakreg}/\texttt{tweakwcs}\footnote{Here, \texttt{tweakwcs} and \texttt{tweakreg} perform the same mathematical transformations.} to perform the transformations: 

 \begin{equ}[ht]\captionsetup{width=.9\linewidth}
\begin{equation}
\centering
\label{eqn:U}
\begin{aligned}
U = P_0 + P_1x' + P_2y'
\end{aligned}
\end{equation}
\begin{equation}
\centering
\label{eqn:V}
\begin{aligned}
V = Q_0 + Q_1x' + Q_2y'
\end{aligned}
\end{equation}
\caption*{Equations 1 and 2: used by \texttt{tweakreg}/\texttt{tweakwcs} to perform the linear transformation between two coordinate systems.} 
\end{equ}

$U$ and $V$ are the positions of the reference catalog (Gaia) as projected into the tangent plane. $x$ and $y$ are the source coordinates measured by \texttt{hst1pass} on the WFC3/UVIS images, and $x'$ and $y'$ are corrected for WFC3/UVIS geometric distortions by applying the current IDCTAB\footnote{The current IDCTAB at the time of publication of this report is \say{2731450pi\textunderscore idc.fits}, delivered in 2016. Please see the HST \href{https://hst-crds.stsci.edu/}{Calibration Reference Data System}. } from the HST \href{https://hst-crds.stsci.edu/}{Calibration Reference Data System} (CRDS) and then projected into the tangent plane using the image's WCS. Equations 1 and 2 above are written as a matrix equation and solved with a standard least squares algorithm. Residual outliers are removed and the least square optimization is repeated a specified number of times.

Parameters $P_0$ and $Q_0$ correspond to the offset between the two systems, i.e. the \say{shift} term of the linear distortion. Parameters $P_1, P_2, Q_1,$ and $Q_2$ are the linear terms between the two catalogs in tangent plane space and can be used to find the scale, rotation, and skew, as follows \citep{2018wfc..rept....9M}:

\begin{equ}[ht]
\captionsetup{width=.9\linewidth}

\begin{equation}
\centering
\label{eqn:x_rot}
\begin{aligned}
Rotation_x = arctan\left( \frac{-P_1}{Q_1} \right)
\end{aligned}
\end{equation}

\begin{equation}
\centering
\label{eqn:y_rot}
\begin{aligned}
Rotation_y = arctan\left(\frac{P_2}{Q_2}\right)
\end{aligned}
\end{equation}

\begin{equation}
\centering
\label{eqn:global_rot}
\begin{aligned}
Rotation_{global} = arctan \left( \frac{P_2 - Q_1}{ P_1+ Q_2} \right)
\end{aligned}
\end{equation}

\caption*{Equations 3, 4 and 5: used by \texttt{tweakreg}/\texttt{tweakwcs} to calculate the rotation angle in the $x-$ direction and rotation angle in the $y-$ direction, as well as the global rotation.} 
\end{equ}

 \begin{equ}[ht]
\captionsetup{width=.9\linewidth}
\begin{equation}
\centering
\label{eqn:skew_glob}
\begin{aligned}
Skew_{global} = Rotation_x - Rotation_y
\end{aligned}
\end{equation}

\caption*{Equation 6: used by \texttt{tweakreg}/\texttt{tweakwcs} to calculate the global skew term, which is the non-orthogonality between the two principle axes.} 
\end{equ}

\begin{equ}[ht]\captionsetup{width=.9\linewidth}
\begin{equation}
\centering
\label{eqn:scalex}
\begin{aligned}
Scale_x =\sqrt{ (P _1^2 + Q_1^2 )}
\end{aligned}
\end{equation}

\begin{equation}
\centering
\label{eqn:scaley}
\begin{aligned}
Scale_y=\sqrt{ (P _2^2 + Q_2^2 )}
\end{aligned}
\end{equation}
\caption*{Equations 7 and 8: used by \texttt{tweakreg}/\texttt{tweakwcs} to calculate the $x-$ and $y-$scale between the two coordinate systems.} 
\end{equ}

It should be noted, however, that these terms are not the \textit{absolute} terms, but rather the relative difference between the initial geometric distortion solution terms for the image, and the solution that best aligns the image to Gaia reference catalog.  These terms are then applied to the image WCS to correct the astrometric solution and the updated corresponding WCS keywords are stored in the FITS header.

\subsubsection{Extracting Transform Components from the WCS}
\label{sec:transform-WCS}

The corresponding terms that are used for the absolute transform (from pixel coordinates to sky coordinates) can be extracted from the WCS keywords.  The absolute rotation between the detector $y$-axis and North (orient), $x$ and $y$ pixel scales (degrees per pixel) and skew terms (absolute angle detector $x-$ and $y-$ axes) can be calculated from the CD matrix keywords of the WCS:

 \begin{equ}[ht]
\captionsetup{width=.9\linewidth}
\begin{equation}
\centering
\label{eqn:scalex_cd}
\begin{aligned}
Scale_x =\sqrt{ CD_{11}{}^2 + CD_{12}{}^2 }
\end{aligned}
\end{equation}

\begin{equation}
\centering
\label{eqn:scaley_cd}
\begin{aligned}
Scale_y =\sqrt{ CD_{21}{}^2 + CD_{22}{}^2 }
\end{aligned}
\end{equation}

\begin{equation}
\centering
\label{eqn:orient_cd}
\begin{aligned}
Orient = arctan \left( \frac{CD_{12}}{CD_{22}} \right)
\end{aligned}
\end{equation}

\begin{equation}
\centering
\label{eqn:skew_cd}
\begin{aligned}
Skew = Orient - arctan \left(\frac{-CD_{11}}{CD_{21}} \right) - 90
\end{aligned}
\end{equation}

\caption*{Equations 9 - 12: Equations describing the extraction of the terms used for the absolute transform (from pixel coordinates to sky coordinates) from the WCS keywords stored in the science extension of the FITS header of each image.} 
\end{equ}

Each $CD_{ij}$ is stored in the WCS of each science FITS header as keyword \verb+CDi_j+.  The shift component of the solution is stored using the \verb+CRPIx+ and \verb+CRVAL+ keywords, which represent the position of the reference pixel in pixel and sky space, respectively.  These equations, while similar to equations 1 and 2, compute absolute transformations to the physical (sky) coordinates.  

\textbf{To measure the time evolution of the linear terms, we compute these absolute transformations for both the initial WCS as generated by just using the IDCTAB, and the updated WCS after the image has been aligned to Gaia.  Comparing the corresponding transformations derived from each of the WCS's allows measurement of the errors in the IDCTAB}.  The calculated values for the orient and skew from the IDCTAB WCS are subtracted from their counterparts from the Gaia-aligned WCS. These differences reflect the error in degrees of absolute rotation in the IDCTAB in the case of orient, and error of angle between the $x$ and $y$ axes in the IDCTAB in the case of skew. The calculated values of $x$-scale and $y$-scale from the Gaia WCS are divided by their counterparts in the IDCTAB WCS. These ratios reflect what the true pixel scale is relative to that calculated in the IDCTAB.  To calculate the shift between the IDCTAB and Gaia frame, we project the \texttt{CRVAL} of the IDCTAB into pixel space using the Gaia WCS, and subtract this projected position from the original \verb+CRPIx+ values.  This reflects how the image was shifted when aligning to the Gaia frame.  While the shift term may represent errors in the IDCTAB, it is likely dominated by errors in the positions of the guide stars used during an observation, or uncertainties in the positions of the Fine Guidance Sensors (FGS) in the HST focal plane.

For every image run through the pipeline, we calculate the absolute linear transformations for both the initial WCS generated with the IDCTAB and the WCS after the image has been aligned to Gaia. We store the shift, rotation, scale, and skew terms for each set of linear transformations, along with the number of matches between the UVIS image and Gaia, the RMSE of the fit to Gaia, and other descriptive metadata in a dataframe. We compare the absolute transformations calculated using the IDCTAB WCS to those using the Gaia-aligned WCS as described in the previous paragraph, and store these difference (or ratio) values in the dataframe, as well.

\subsubsection{Clipping the Data}
\label{sec:clipping}

Once the images and their transformations are collated into one large dataframe, we cut out any poorly fit images from the dataset. Images with only a small number of matches to Gaia or images of crowded fields could result in a poor fit. Figure 1 plots the RMSE (root mean square error) of each image's fit to Gaia vs. the number of matches to Gaia. We choose to cut the dataset to include only images with an RMSE less than 0.2 pixels, indicating that the fit to Gaia is good, reducing the uncertainty in the linear terms of the fit. We cut the initial set of 9,700 images to a set of 7,491 well-fit images which we use for for our subsequent analysis.

\input{plots/Gaia_RMSE_plot}

Next, we perform two iterations of sigma clipping on the data to within three standard deviations of the mean. Sigma clipping is performed on the difference (or ratio, in the case of the scale term) values between each term in the IDCTAB linear transformations and the corresponding term in the Gaia-aligned linear transformations. When analyzing and plotting these terms by filter, we perform the sigma clipping for each filter independently. When reporting the statistics for all data, we perform sigma clipping on the entire dataset at once. We then fit a line to each computed quantity vs time using Scipy's \texttt{linregress()} routine \citep{2020SciPy-NMeth}, which performs a linear least-squares regression.

\section{Results}
\label{sec:results}

We evaluate how WFC3/UVIS geometric distortion has evolved over time with respect to the Gaia catalog by looking at each of the terms (shift, rotation, scale, and skew) individually and computing the offsets for each term between the IDCTAB transformations and the Gaia-aligned transforms. We provide plots of these offsets over time, and report the maximum possible change, in pixels, that the evolution of each linear term would cause. Because a change to the geometric distortion would cause the greatest change at the edges of the detector, we report the change, in pixels, at the edge of the detector for each term. 

\textbf{Note that, for the discussion of each term, we are looking at the relative difference between the WFC3/UVIS pipeline linear transformations\footnote{The pipeline linear transformations are from a combination of the WFC3/UVIS geometric distortion reference files: IDCTAB, NPOLFILE, and D2IMFILE.} and the Gaia-aligned linear transformations, rather than the absolute value of the shift, rotation, scale, or skew terms.}

\subsection{Relative Shift Offsets}
\label{sec:shift}

By far the largest term of the geometric distortion solution with respect to Gaia is the shift term. This is the 0th order constant term, in $x$ and in $y$, that lines up the WFC3/UVIS image coordinates with the standard Gaia coordinates. In other words, the WFC3/UVIS sources must be shifted by some amount $dx$ and $dy$ to match the Gaia source positions. Tables \ref{tab:xshift} and \ref{tab:yshift} provide a summary of $x$ and $y$ shift offsets and Figures \ref{fig:shift_overplot} and \ref{fig:shift_subplot} show the relative shifts between WFC3/UVIS and Gaia over time.

\input{tables/shift_tables}

Fitting lines to the $x$ and $y$ relative shifts from WFC3/UVIS to Gaia source positions, some evolution over time is apparent, as seen in Figures \ref{fig:shift_overplot} and \ref{fig:shift_subplot} . The maximum change (calculated for all filters together) over 13 years is about $2 \pm  5$ pixels in $x$ (0.17 pixels/year) and $-4 \pm 5$ pixels in $y$ (-0.27 pixels/year). This change is small compared to the uncertainty in pointing within any given observation, as demonstrated by the spread of the shift values. The uncertainty improves to about $\pm 3$ pixels in $x$ and $\pm 4$ pixels in $y$ after October 2017, when the Gaia astrometric catalogs were used to update the HST Guide Star Catalogs (GSC v2.40) and improve the accuracy of HST absolute astrometry \citep{hoffmann2021drizzlepac}. The large scatter in the shift offsets to the Gaia catalog is largely due to sources of error outside of the distortion calibration, including uncertainty in telescope pointing, as mentioned in the Discussion section of this report. 

\input{plots/shift_plots}
\clearpage

\subsection{Relative Rotation Offsets}
\label{sec:rotation}
The next term we evaluate is the rotation term. This term describes how the WFC3/UVIS coordinate system is rotated with respect to Gaia. In Figure \ref{fig:orient_overplot}, we plot this evolution in degrees, as well as in pixels, as a function of time for five filters; Figure \ref{fig:orient_subplot} separates the data by filter.  The maximum change in pixels (at the edge of the detector) is calculated as the distance from the center of the FOV (the axis of rotation) to the edge of the detector multiplied by the sine of the difference in rotation angle, as described by Equation 13.

 \begin{equ}[ht]
\captionsetup{width=.9\linewidth}
\begin{equation}
\centering
\label{eqn:rot_to_pix}
\begin{aligned}
2048 \times sin(\Delta\theta)
\end{aligned}
\end{equation}
\caption*{Equation 13: used to calculate the maximum change in pixels at the edge of the detector caused by the relative rotation offset between the WFC3/UVIS frame and the Gaia DR3 frame.} 
\end{equ}

As seen in Figures \ref{fig:orient_overplot} and \ref{fig:orient_subplot}, the temporal evolution of the relative rotation offset with regard to Gaia is not well-described by a simple linear fit. A linear least squares regression  over all filters yields a slope of -0.00011 degrees per year ($-0.05 \pm 0.22$ pixels at the edge of the detector over 13 years). The uncertainty in the rotation offset is significantly larger than the calculated linear rate of change in the rotation over time reported in Table \ref{tab:rotation}. \textbf{ We find a small, stable offset to Gaia of $\mathbf{0.003 \pm 0.004}$ degrees between 2009-2017, with a systematic increase to $\mathbf{0.009 \pm 0.005}$ degrees from 2017-2020 and then a systematic decrease to $\mathbf{-0.003  \pm 0.008}$  degrees from 2020-2022.}  We note that the large rotation offset starting in mid-2017 corresponds to a period of noticeably increased jitter in HST's pointing stability reported by Anderson and Sabbi \citep{2018wfc..rept....7A}. 

\input{tables/rotation_table}
\input{plots/rotation_plots}
\clearpage

\subsection{Relative Scale Offsets}
\label{sec:scale}

To analyze how the WFC3/UVIS scale has evolved over time, we compute the ratio of the Gaia $x-$scale (or $y-$scale) to the WFC3/UVIS $x-$scale (or $y-$scale), as provided in the IDCTAB reference file (see Figures \ref{fig:scale_overplot} and \ref{fig:scale_subplot}, in which a ratio of $1.0$ indicates no residual scale). To assess the greatest possible effect at the edge of the detector, in pixels, that any change in scale would have over 13 years, we compute the slope of the line of best fit for each filter (and for all data) and multiply it by the size, in pixels, of the detector \footnote{The WFC3/UVIS detector size is  $4096 \times 4096$ pixels.}, as described by Equation 14.  

 \begin{equ}[ht]
\captionsetup{width=.9\linewidth}
\begin{equation}
\centering
\label{eqn:scale_to_pix}
\begin{aligned}
4096 \times \frac{Scale_{Gaia}}{Scale_{UVIS}} - 4096
\end{aligned}
\end{equation}
\caption*{Equation 14: used to calculate the maximum change in pixels at the edge of the detector caused by the relative scale ratio between the WFC3/UVIS frame and the Gaia DR3 frame.} 
\end{equ}  

An analysis of the WFC3/UVIS $x-$ and $y-$scale terms with respect to Gaia reveals the importance of fitting a scale term when aligning images with DrizzlePac's \texttt{tweakreg} algorithm. As with the other terms in the linear portion of the WFC3/UVIS geometric distortion solution (shift, rotation, and skew), we see a small time evolution, in this case on the order of $0.15 \pm 0.26$ pixels in $x$ and $0.11 \pm 0.27$ pixels in $y$, at the edge of the detector, over 13 years. When analyzing all filters together, the RMS of the scale measurements is greater than the change in scale in this time range, indicating that this change is negligible compared to the uncertainty in the scale term at any given time. However, the scatter (RMS) does not outweigh the temporal scale change in all filters when measured individually. We note that the scatter about the trend line is largest in F275W, and in all other filters the temporal change is not outweighed by the scatter. 

Besides a small change in the WFC3/UVIS $x-$ and $y-$scale terms over time of about $0.2$ pixels, there is a clear systematic offset at all epochs in both the $x-$ and $y-$scale values between filters of up to $0.3$ pixels at the edges of the chip, for instance, between F275W and F814W (see Figure \ref{fig:scale_overplot} and Tables \ref{tab:x-scale} and \ref{tab:y-scale}). This is likely the result of filter-specific geometric distortions which are not accounted for in the current set of reference files. The presence of this filter dependence once again reinforces the need for observers to align their data to ensure astrometric accuracy. 
\input{plots/scale_plots}

\clearpage

\input{tables/scale_tables}

\clearpage

\subsection{Relative Skew Offsets}
\label{sec:skew}
The last linear term that we analyze is the skew term. The skew describes the total amount of non-orthogonality (difference from 90 degrees) between the principal $x$- \& $y$-axes. It is reported as an angle (in degrees), and we compute the maximum change in pixels at the edge of the detector due to a difference in skew angle by multiplying the length of the detector (4096 pixels) by the sine of the difference in skew angle, as described by Equation 15. Those values are presented in Table \ref{tab:skew}. 

 \begin{equ}[ht]
\captionsetup{width=.9\linewidth}
\begin{equation}
\centering
\label{eqn:skew_to_pix}
\begin{aligned}
4096\times sin(\theta)
\end{aligned}
\end{equation}
\caption*{Equation 15: used to calculate the maximum change in pixels at the edge of the detector caused by the relative skew offset between the WFC3/UVIS frame and the Gaia DR3 frame.} 
\end{equ}

\input{plots/skew_plots}

The skew term evolves by about $0.2 \pm 0.3$ pixels over 13 years (a rate of $0.014$ pixels/year or $0.0002$ degrees/year over all filters). Though the RMS varies by filter, the calculated temporal change is generally smaller than or comparable to the RMS (with the exception of F814W), and there are no clear filter-dependent offsets for the skew term (see Figures \ref{fig:skew_overplot} and \ref{fig:skew_subplot}). 

\clearpage
 \input{tables/skew_table}

\section{Discussion}
\label{sec:discussion}

 In order to obtain accurate astrometry from WFC3 images and precise proper motions, images must be aligned to correct for any distortion and offsets. Uncertainty in the telescope pointing is the most prevalent source of offsets, as demonstrated by the large scatter in the $x$ and $y$ shift terms.   This uncertainty in pointing is expected due to errors in the Guide Star Catalogue (GSC) coordinates, which were originally created using photographic plates taken from ground-based telescopes, which have lower centroid precision than required for HST astrometry
 \citep{1990AJ.....99.2019L}. The uncertainties of the GSC positions can also cause small errors in the calculated rotation angle (orient) of the observatory (as two guide stars are often used to constrain the pointing and roll angle).   Proper motions between the photographic plate epoch and HST observations and uncertainties in the positions of WFC3  in HST's focal plane relative to the Fine Guidance Sensors (FGS) also add uncertainty to the absolute astrometry \citep{hoffmann2021drizzlepac}.

The significant scatter observed around the trends for each of the linear terms also may result from positional inaccuracies in either the image catalog or the Gaia catalog. Errors in the positions from either catalog would propagate into the calculation of the linear terms in the astrometric alignment. If these errors were predominantly random, the scatter around the trend line would be expected to decrease by $N^{-1/2}$  as the number of matched stars (N) increases. While the residuals decrease somewhat as N increases, they do not decrease significantly after about 100 matches to Gaia. 

For example, in Figure \ref{fig:skew_residuals} below, the spread of the skew residuals above 100 matches is $\pm 0.065$ pixels ($\pm 0.0009$ degrees) and only decreases to $\pm 0.057$ pixels ($\pm 0.0008$ degrees) above 1500 matches. This suggests the presence of a systematic error in the centroid measurements, or that the observed variation around the trend lines may reflect genuine astrometric deviations.  However, the median error $\left( \frac{ RMS}{\sqrt{N}}\right)$ between image catalog and Gaia catalog positions after alignment is 0.006 pixels, significantly less than the scatter about the trend lines of each of the linear terms.  Thus, the astrometric deviation does not appear to be an artifact of the source positions or transform calculation, indicating real variation. The residuals for the other linear terms follow a similar trend.

\begin{figure}[!ht]
    \centering
    \includegraphics[width=18cm]{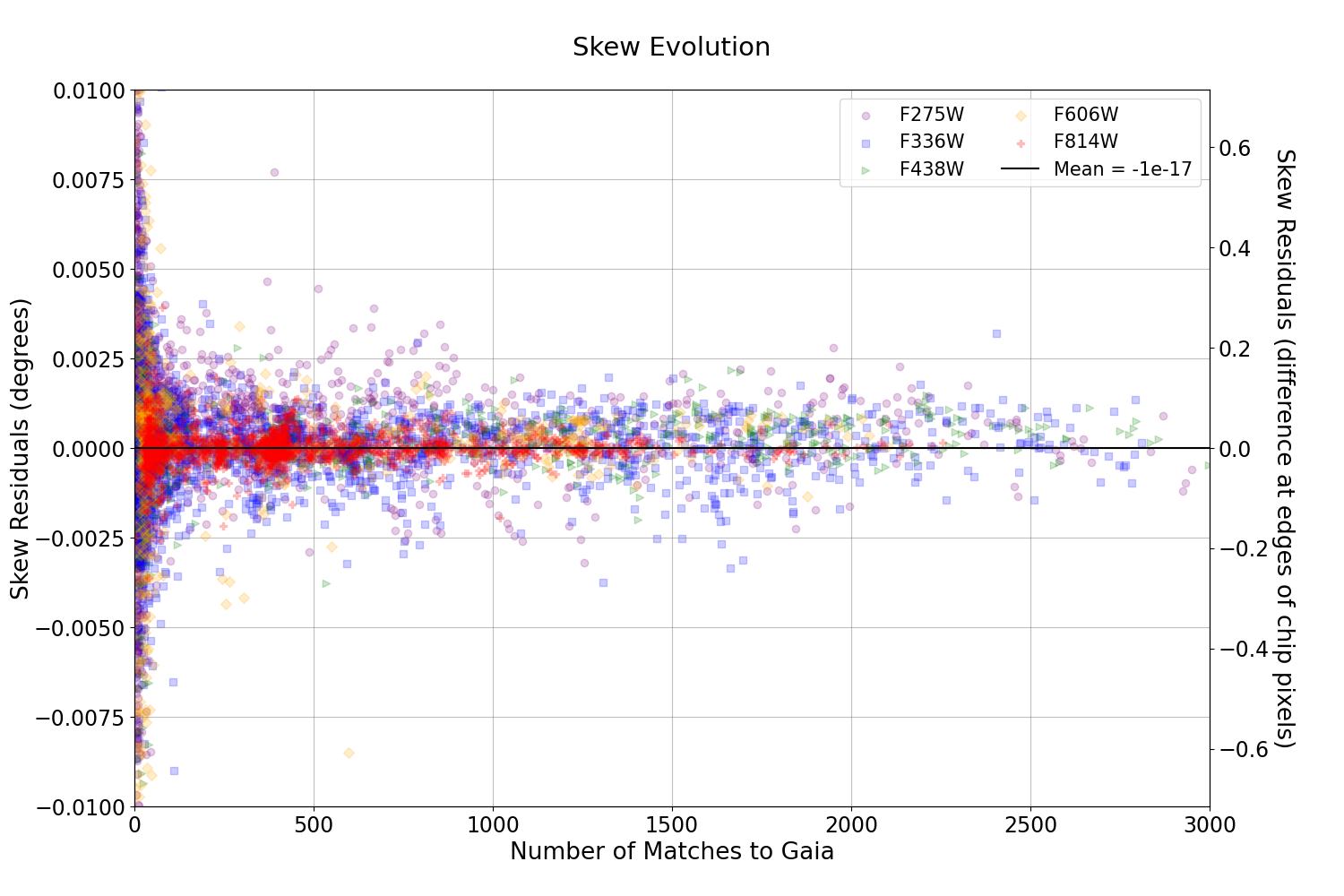}\caption    {Skew residuals vs number of Gaia matches (all filters).}
    \label{fig:skew_residuals}
\end{figure}

The uncertainty in the telescope pointing for any given observation requires all images to be aligned as part of standard image processing procedures, not only to correct for offsets and improve precision astrometry, but also to allow correct processing and image combination (i.e. with DrizzlePac). In the next section, we present the recommended image alignment keywords to correct for any uncertainty and/or temporal variation in the linear terms.

\clearpage
\section{Recommendations for the User}
\label{sec:recommendations}

For the linear terms (shift, rotation, scale, and skew), the temporal evolution of the relative offsets between the IDCTAB and Gaia is primarily dominated by scatter. As such, we recommend that observers continue to align individual observations using \texttt{tweakreg} (or \texttt{tweakwcs}) in DrizzlePac, as is currently recommended. We further recommend that observers fit a scale term (and possibly a skew term) during image alignment using one of two fit geometry keywords, depending on the number of source matches, as described in the following paragraphs. 

Aligning images to each other will correct for any relative astrometric errors in the shift, rotation, scale, or skew terms. While not always possible due to low source density, aligning images to Gaia will correct for relative and absolute astrometric errors. In both cases, the scale offset between filters will be corrected by alignment and thus alignment is necessary if precision astrometry between filters is required. 

The calibration pipeline processing for data in the Mikulski Archive for Space Telescopes (MAST) includes a step in which astrometric alignment to Gaia is attempted for WFC3 images\footnote{See readthedocs in DrizzlePac for a description of the \href{https://drizzlepac.readthedocs.io/en/latest/mast_data_products/pipeline-astrometric-calibration-description.html}{Pipeline Astrometric Calibration.}}. In the case where this alignment attempt is unsuccessful, the software will use an \textit{'a priori'} WCS solution, which simply recalculates the WCS with more accurate guide star positions but does not perform any alignment. These solutions will have the header keyword (\texttt{WCSNAME} set to \verb+IDC_2731450pi-GSC240+ or \verb+IDC_2731450pi-HSC30+). In this case, the offsets to Gaia are smaller, yet still significant enough to require realignment for high precision astrometry.  Additionally the scale offset between filters remains, as the pixel scale is unchanged in \textit{a priori} solutions.   

In the case where the Gaia alignment is successful, the data will use the resulting aligned WCS, called an \textit{'a posteriori'} solution.  The \texttt{WCSNAME} keyword will take the form\\ \verb+FIT_<method>_<catalog>+, where \verb+catalog+ will be \verb+GAIAeDR3+, \verb+GAIADR2+, or \verb+GAIADR1+. When there are not enough matched sources for a successful Gaia alignment, alternate reference catalogs may used, including \verb+2MASS+ or \verb+GSC242+, the Gaia-supplemented Guide Star Catalog.  While this alignment generally improves absolute astrometry, the precision of the alignment may be insufficient in fields with few Gaia sources.  Furthermore, in some cases, performing the alignment to Gaia may slightly degrade the relative astrometry between exposures in the same filter or in different filters. 

If the \verb+<method>+ portion of the \texttt{WCSNAME} is \verb+REL+ or \verb+SVM+, then the relative astrometric offsets have likely been removed, though the absolute astrometric error may still remain due to lower precision in these catalogs (compared to Gaia).  In the case where there are multiple different catalogs used for a given set of images, the differences between the catalogs will likely result in poorer relative astrometry.  For more details, see \citep{2022wfc..rept....6M}.

If precision relative astrometry is desired, users may need to realign their data.  To decide, check the the value of the \texttt{NMATCHES} keyword, which records the number of matches to Gaia used in the MAST alignment. \textbf{If the value of the \textbf{\texttt{NMATCHES}} keyword is above 20 matches, it may not be necessary to realign data, as the alignment in the MAST pipeline should correct any relative offsets as well as absolute offsets}. If \texttt{NMATCHES} indicates fewer than 20 matches, consider resetting the WCS to the original \textit{'a priori'} solution and realigning the data, as described in the DrizzlePac notebook \href{https://spacetelescope.github.io/hst_notebooks/notebooks/DrizzlePac/using_updated_astrometry_solutions/using_updated_astrometry_solutions.html}{Improving Astrometry Using Alternate WCS Solutions.}

\subsection{Using DrizzlePac's \texttt{tweakreg} Task}
\label{sec:tweakreg}
 \href{https://drizzlepac.readthedocs.io/en/latest/}{DrizzlePac} \citep{drizzlepac} is a software package for aligning and combining HST images in a process known as \say{drizzling}. Combining images using AstroDrizzle (DrizzlePac's primary user interface) requires that the WCS information in the headers of each input image align to within sub-pixel accuracy. The \texttt{tweakreg} task allows the user to align sets of images to each other and/or to an external astrometric reference frame or image. 

 Our recommendation to users as it relates to uncertainty and offsets in the linear terms of the image WCS (World Coordinate System) is to fit a shift, rotation, scale, and possibly a skew term while aligning images. To fit a \textbf{scale term} in \texttt{tweakreg}, one must set the parameter \texttt{fitgeometry} to  \texttt{`rscale'} (which also fits shift and rotation terms), as seen in Code Example \ref{code}. For those performing high-precision astrometric calculations with WFC3/UVIS data across long baselines, we recommend fitting a \textbf{skew term} when aligning images in \texttt{tweakreg}. To do so, set \texttt{fitgeometry} to \texttt{`general'}, which calculates skew terms along with shift, rotation, and x and y scales by allowing an independent scale and rotation for each axis. 
\renewcommand{\lstlistingname}{Code Example}

 \begin{minipage}[!ht]{0.9\linewidth} \captionsetup{width=.9\linewidth} \vspace{2ex}
\begin{lstlisting}[
frame=tb, 
caption=An example of how to call DrizzlePac's \texttt{tweakreg} routine to align individual exposures., 
captionpos=b, belowcaptionskip=0.2cm, 
label=code, 
language=python
]
# Source Detection parameters
cw = 3.5               # ~2x PSF FWHM (UVIS= 3.5 pix, IR= 2.5 pix)
thresh = 500           # Threshold (UVIS= 500-2000, IR= 5-100)

# Reference Catalog parameters
refcat = 'gaia.cat'    # User provided Gaia catalog
wcsname = 'Gaia_DR3'   # Name of new WCS solution 

# Fitting parameters
fitgeom = 'rscale'     # Fit geometry ('rscale', 'general')  
min_obj = 20           # Number of matches for a successful fit 
radius = 0.1           # Search radius in arcsec      

from drizzlepac import tweakreg
tweakreg.TweakReg('*flc.fits',
                  imagefindcfg={'threshold': thresh, 'conv_width': cw},
                  refcat = refcat,
                  wcsname = wcsname,
                  shiftfile = True,
                  outshifts = 'shifts.txt',
                  fitgeometry = fitgeom,
                  minobj = min_obj,
                  searchrad = radius,
                  interactive = False,       
                  see2dplot = False,
                  updatehdr = False)
\end{lstlisting}

\end{minipage}
 
The code snippet above is excerpted from the Jupyter notebook \say{\href{https://spacetelescope.github.io/hst_notebooks/notebooks/DrizzlePac/align_to_catalogs/align_to_catalogs.html}{Aligning HST images to an Absolute Reference Catalog}}. For more details on aligning to Gaia, see notebook Sections 3.3 through 3.6 which describe how to inspect the fitting results, including the output shift file (\verb|shifts.txt|) and residual plots (\verb|residuals_*.png|). Once the user is satisfied with the fitting results, \verb|tweakreg| can be rerun a second time with the parameter \verb|updatehdr| set to \verb|True|. Please refer to this notebook for an in-depth walkthrough of aligning HST images to an external catalog and more information about other \texttt{tweakreg} keywords.

In the case where only relative alignment is desired (i.e. images only need to be aligned to each other, rather than 
 to an external catalog), the \verb+refcat+ keyword argument can be set to \verb+None+ (see the Jupyter notebook \say{\href{https://spacetelescope.github.io/hst_notebooks/notebooks/DrizzlePac/align_multiple_visits/align_multiple_visits.html}{Optimizing Image Alignment for Multiple HST Visits}} for more information). In this case, \texttt{tweakreg} will use the first input image as a reference image to which it will align the other input images. 
  
\clearpage 
\section{Conclusions and Future Work}
\label{sec:Conclusions}

After assessing the temporal evolution of the WFC3/UVIS linear geometric distortion terms using 13 years of data and 7,491 images of a large variety of targets, we reaffirm the need for observers to correct for any residuals in the linear terms (from HST to Gaia or between different HST filters) by realigning their FLT/FLC images with \texttt{tweakreg}. We note that, in some cases, the alignment performed in the MAST calibration pipeline may be sufficient to correct for any relative and absolute offsets in the linear geometric distortion terms. The observed time dependence in the linear distortion terms is generally outweighed by the uncertainty in those terms for any given observation, especially between 2018-2022 when the rotation residuals can be as large as $\pm 0.01$ degrees. Additionally, we find a larger filter-dependent offset in the scale term of up to 0.3 pixels between filters at all epochs; therefore, alignment would likely still be required even if a time-dependent IDCTAB reference file were available to account for the slow change in scale and skew terms over time. Out key results are as follows:

\begin{itemize}
    \item In all linear terms of the WFC3/UVIS geometric distortion solution there is a small temporal evolution with respect to Gaia DR3. 
    \item The WFC3/UVIS scale term exhibits a filter-dependent offset of up to 0.3 pixels between filters.
    \item The WFC3/UVIS relative rotation offset residuals increased between 2018-2022 to up to $\pm 0.01$ degrees.
    \item We encourage users to inspect the MAST pipeline alignment, which may be sufficient to correct for any relative and absolute offsets in the linear geometric distortion terms.
    \item We encourage users to manually re-align their FLT/FLC images using \texttt{tweakreg} as necessary. For images with very few stellar sources, users can reset the primary WCS to use the \textit{\say{a priori}} WCS solution which includes distortion correction but no additional fitting. 
\end{itemize}

Future reports will examine the potential implementation of time-dependent geometric distortion solutions in the WFC3/UVIS IDCTAB, as well as a global scale correction for all filters with respect to Gaia DR3.  We will also investigate whether the large rotation offsets to Gaia are still present at the current epoch. Ultimately, as long as observers continue to manually align WFC3/UVIS images and use the recommended fitting geometry in \texttt{tweakreg} as part of their standard image processing (or determine the MAST alignment to be sufficiently good), any errors in the linear distortion terms will be mitigated.

\pagebreak 
\section{Acknowledgements}

We thank Clare Shanahan for the development of the algorithm used to refit the linear terms of the astrometric transform.  We also thank the WFC3 Astrometric Calibration group for their input and review of this document.  We also thank Jennifer Mack and Mariarosa Marinelli for their review and edits.

This work has made use of data from the European Space Agency (ESA) mission
{\it Gaia} (\url{https://www.cosmos.esa.int/gaia}), processed by the {\it Gaia}
Data Processing and Analysis Consortium (DPAC,
\url{https://www.cosmos.esa.int/web/gaia/dpac/consortium}). Funding for the DPAC
has been provided by national institutions, in particular the institutions
participating in the {\it Gaia} Multilateral Agreement.

\bibliography{ref}
\bibliographystyle{aasjournal}
\nocite{*}

\end{document}

%% file: tables/Data_by_filter.tex
\begin{minipage}[!b]{0.9\linewidth} \vspace{3ex}
\normalsize
    \centering 
    \def\arraystretch{1.3} 
    \begin{tabular}{|c|c|} \hline
\textbf{Filter}  &  \textbf{Number of Images}  \\ \hline
         All&   7,491  \\ \hline 
         F275W&   1,134  \\ \hline 
         F336W&   1,945  \\ \hline 
         F438W&    572 \\ \hline 
         F606W&    1,381 \\ \hline 
         F814W&    2,598 \\ \hline

    \end{tabular}
    
\captionsetup{type=table}
\captionof{table}{The number of images, by filter, collected for analysis of the temporal evolution of the linear terms in the WFC3/UVIS geometric distortion solution with respect to the Gaia DR3 catalog.}
    \label{tab:num_images_by_filter}
\vspace{2ex}
\end{minipage}

%% file: plots/Gaia_RMSE_plot.tex
\begin{figure}[!ht]
    \centering
    \includegraphics[width=18cm]{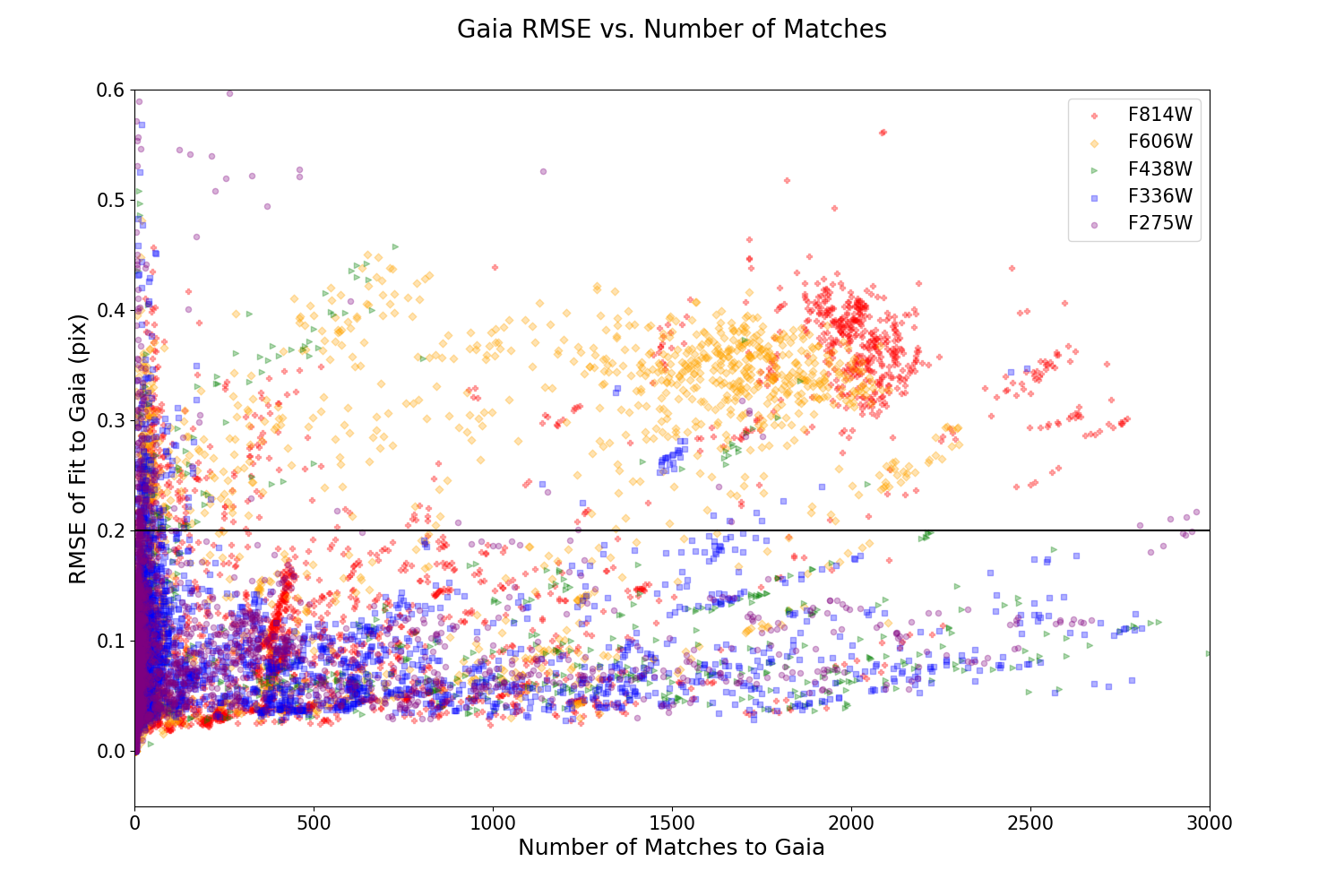}
    \caption{RMSE of each image's fit to Gaia DR3 vs. the number of matches. A smaller RMSE indicates a more accurate fit. We cut out any images with an RMSE above 0.2.}
    \label{fig:gaia_rms_plot}
\end{figure}

%% file: tables/shift_tables.tex
\newcommand{\specialcell}[2][c]{%
  \begin{tabular}[#1]{@{}c@{}}#2\end{tabular}}

\begin{minipage}[!ht]{1\linewidth} 
\captionsetup{width=.9\linewidth}  \vspace{2ex}
\normalsize
    \centering 
    \def\arraystretch{1.3} 
    \begin{tabular}{|c|c|c|c|} \hline

\textbf{Filter} &  \specialcell{\textbf{X-Shift: Average Change} \\ (pix)}  &  \specialcell{\textbf{X-shift: RMS} \\ (pix) } &  \specialcell{\textbf{X-shift: Slope} \\ ($\frac{pix}{year}$) }  \\ \hline
         All&  2&  +/- 5&0.17\\ \hline 
         F275W&  2&  +/- 4&0.15\\ \hline 
         F336W&  2&  +/- 5&0.17\\ \hline 
         F438W&  2&  +/- 5& 0.18\\ \hline 
         F606W&   4&  +/- 6&0.30\\ \hline 
         F814W&  2&  +/- 4&0.14\\ \hline

    \end{tabular}
    
\captionsetup{type=table}
\captionof{table}{X-shift offset, WFC3/UVIS to Gaia DR3. The maximum difference in pixels at edge of detector from 2009 to 2022, the RMS of this difference in pixels, and the slope of the line of best fit in pixels/year are reported.}
    \label{tab:xshift}
\vspace{2ex}
\end{minipage}

\begin{minipage}[!h]{1\linewidth} \captionsetup{width=.9\linewidth}  \vspace{2ex}
\normalsize
    \centering 
    \def\arraystretch{1.3} 
    \begin{tabular}{|c|c|c|c|} \hline
\textbf{Filter}  &  \specialcell{\textbf{Y-Shift: Average Change} \\ (pix)}  &  \specialcell{\textbf{Y-shift: RMS} \\ (pix) } &  \specialcell{\textbf{Y-shift: Slope} \\ ($\frac{pix}{year})$ }  \\ \hline
         All&  -4&  +/- 5&-0.27\\ \hline 
         F275W&  -2&  +/- 5&-0.16\\ \hline 
         F336W&  -2&  +/- 5&-0.18\\ \hline 
         F438W&  -5&  +/- 6&-0.41\\ \hline 
         F606W&  -6&  +/- 7&-0.44\\ \hline 
         F814W&  -4&  +/- 5&-0.33\\ \hline

    \end{tabular}
    
\captionsetup{type=table}
\captionof{table}{Y-shift offset, WFC3/UVIS to Gaia DR3. The maximum difference in pixels at edge of detector from 2009 to 2022, the RMS of this difference in pixels, and the slope of the line of best fit in pixels/year are reported.}
    \label{tab:yshift}
\vspace{2ex}
\end{minipage}

%% file: plots/shift_plots.tex
\begin{figure}[!ht]
    \includegraphics[width=16cm]{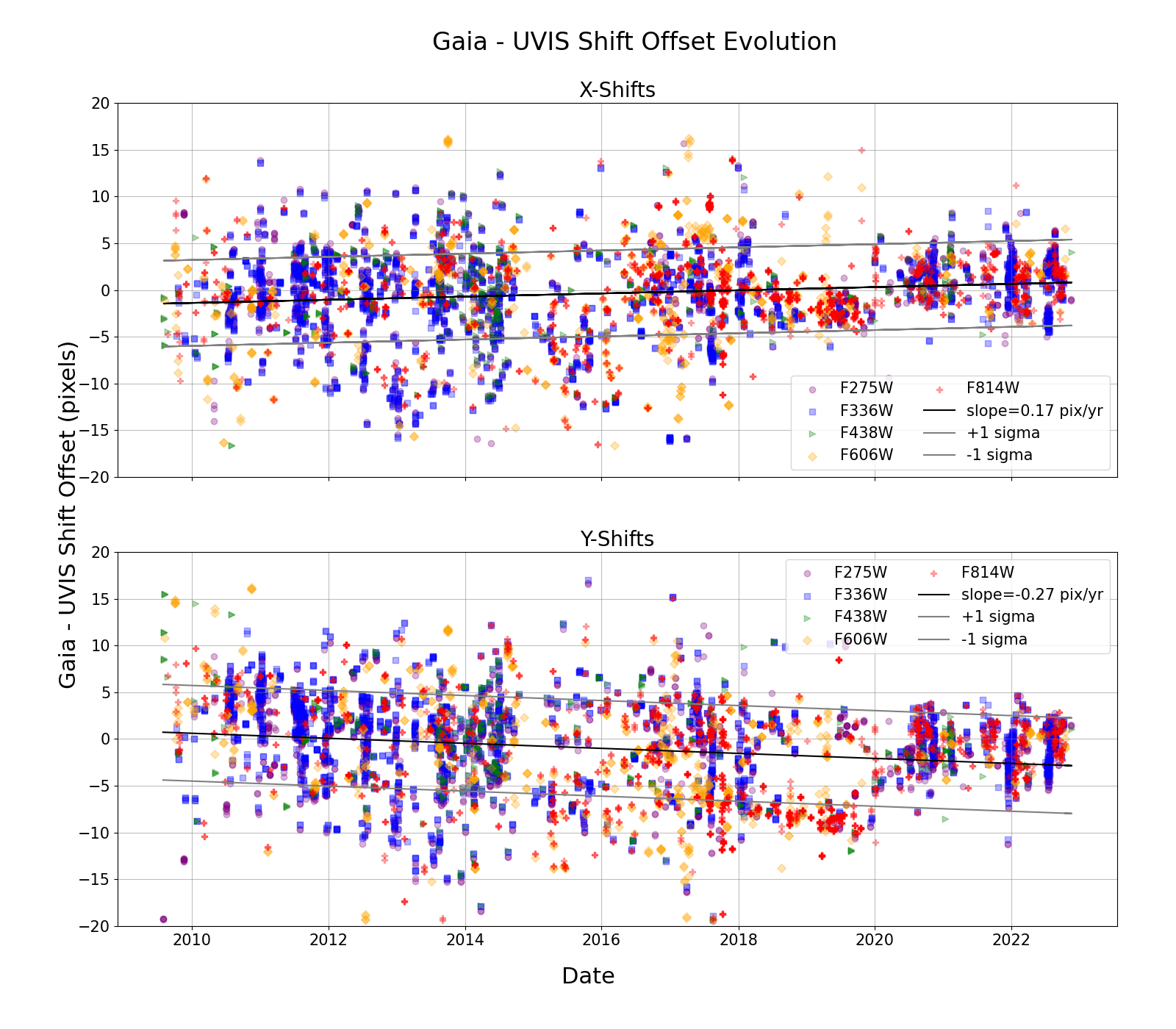}
    \caption{Shift term offsets, WFC3/UVIS to Gaia DR3. Each color and shape combination represents a separate filter. A line of best fit is plotted in black.}
    \label{fig:shift_overplot}
\end{figure}

\begin{figure}[!ht]
    \centering
    \includegraphics[width=16cm]{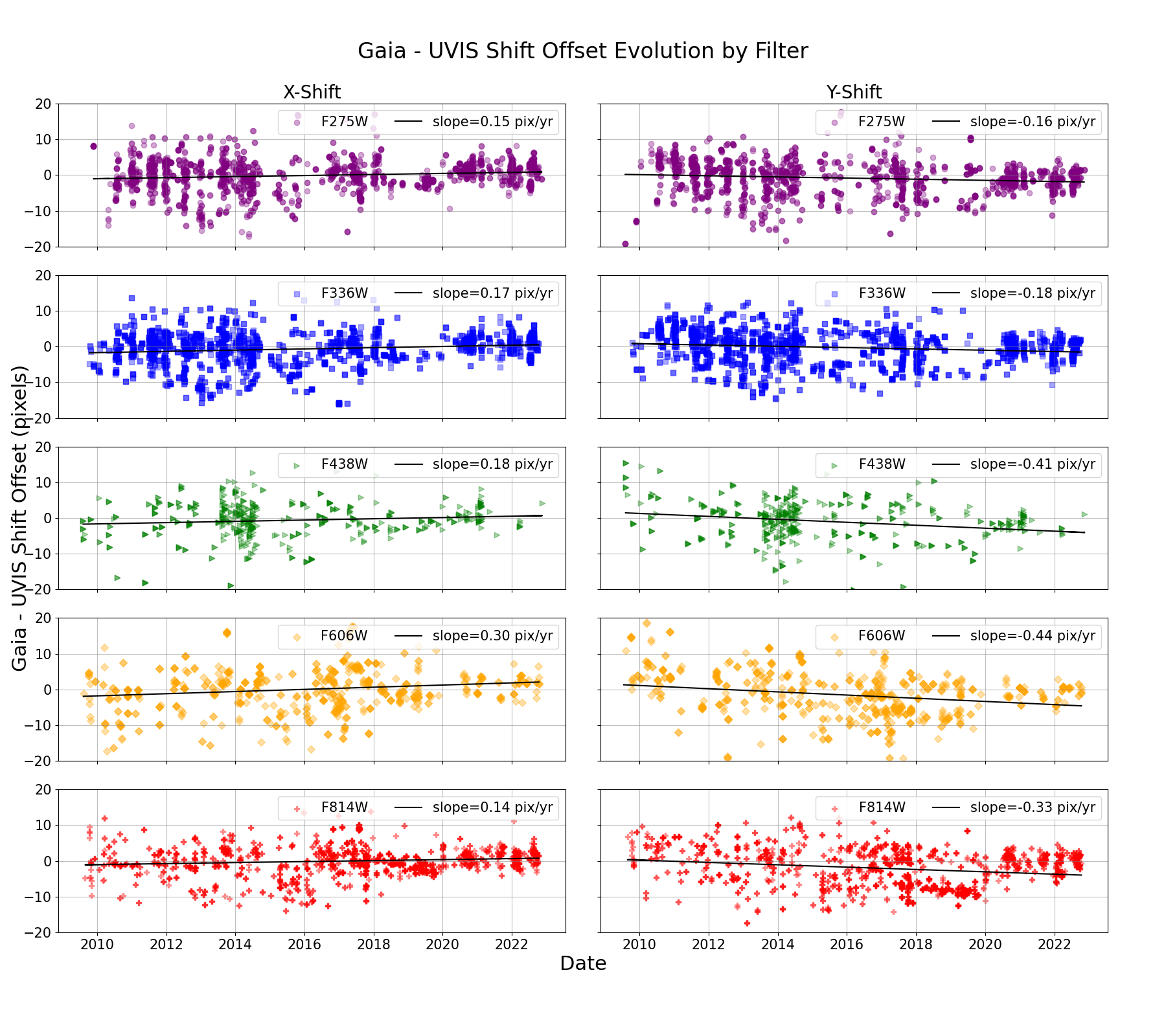}
    \caption{Shift term offsets, WFC3/UVIS to Gaia DR3, by filter.}
    \label{fig:shift_subplot}
\end{figure}

%% file: tables/rotation_table.tex
\begin{minipage}[!ht]{1\linewidth} \captionsetup{width=.9\linewidth}  \vspace{2ex}
\normalsize
    \centering 
    \def\arraystretch{1.3} 
    \begin{tabular}{|c|c|c|c|} \hline
\textbf{Filter}  &  \specialcell{\textbf{Rotation: Average Change} \\ (pix)}  &  \specialcell{\textbf{Rotation: RMS} \\ (pix) } &  \specialcell{\textbf{Rotation: Slope} \\ ($\frac{deg}{year})$ }  \\ \hline
         All&   -0.05&    +/- 0.2&-1E-4\\ \hline 
         F275W&   -0.1&    +/- 0.3 &-3E-4\\ \hline 
         F336W&   -0.1&    +/- 0.2 &-2E-4\\ \hline 
         F438W&    -0.1&    +/- 0.2 &-1E-4\\ \hline 
         F606W&    -0.0003&    +/- 0.2 &-5E-7\\ \hline 
         F814W&    0.005&    +/- 0.2 &1E-5\\ \hline

    \end{tabular}
    
\captionsetup{type=table}
\captionof{table}{Rotation offset, WFC3/UVIS to Gaia DR3. The maximum difference in pixels at edge of detector due to rotation offsets from 2009 to 2022, the RMS of this difference in pixels, and the slope of the line of best fit in degrees/year are reported.}
    \label{tab:rotation}
\vspace{2ex}
\end{minipage}

%% file: plots/rotation_plots.tex
\begin{figure}[!b]
    \centering
    \includegraphics[width=16cm]{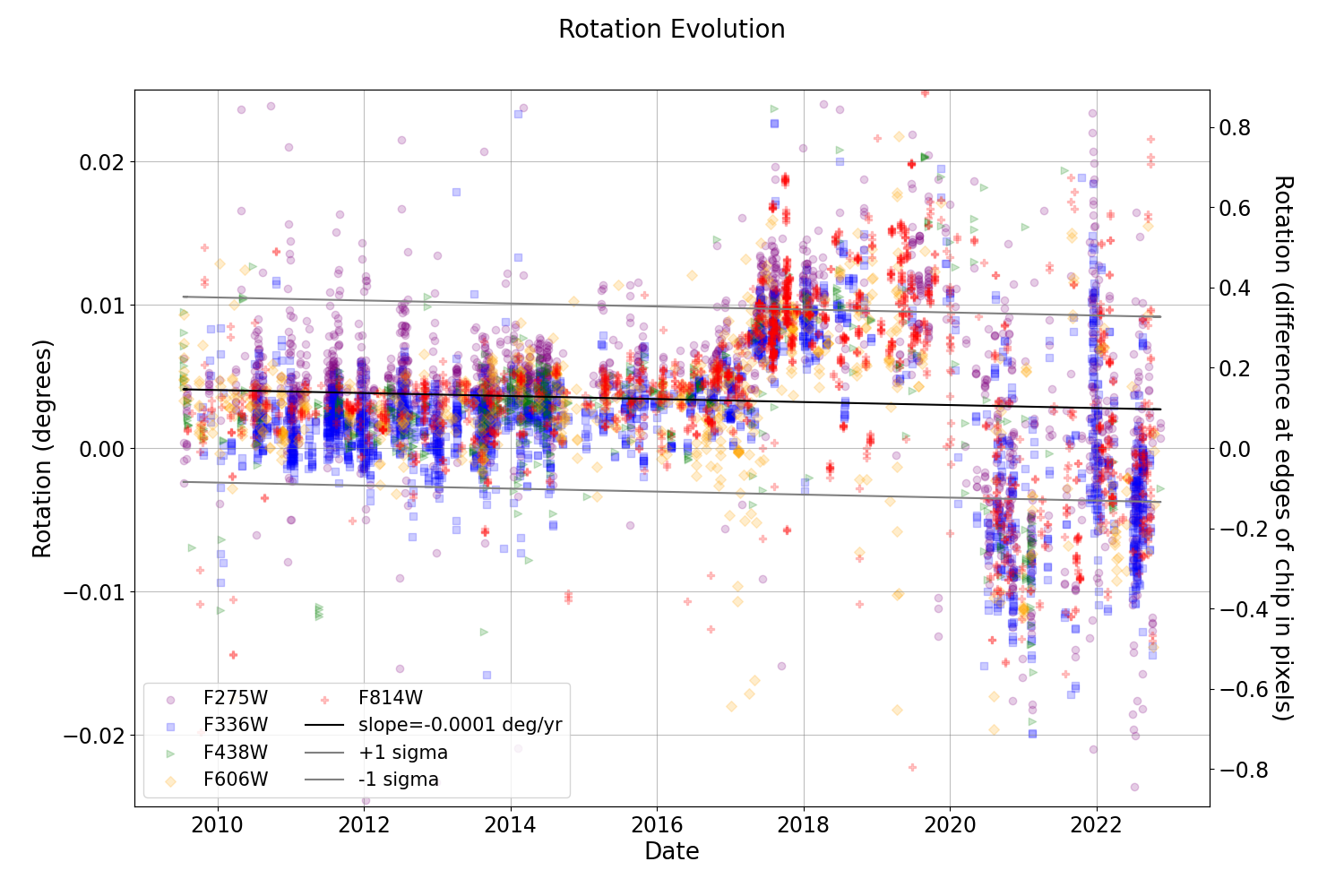}
    \caption{Rotation term offsets: WFC3/UVIS to Gaia DR3. Each color and shape combination represents a separate filter. A line of best fit is plotted in black.}
    \label{fig:orient_overplot}
\end{figure}

\begin{figure}[!ht]
    \centering
    \includegraphics[width=16cm]{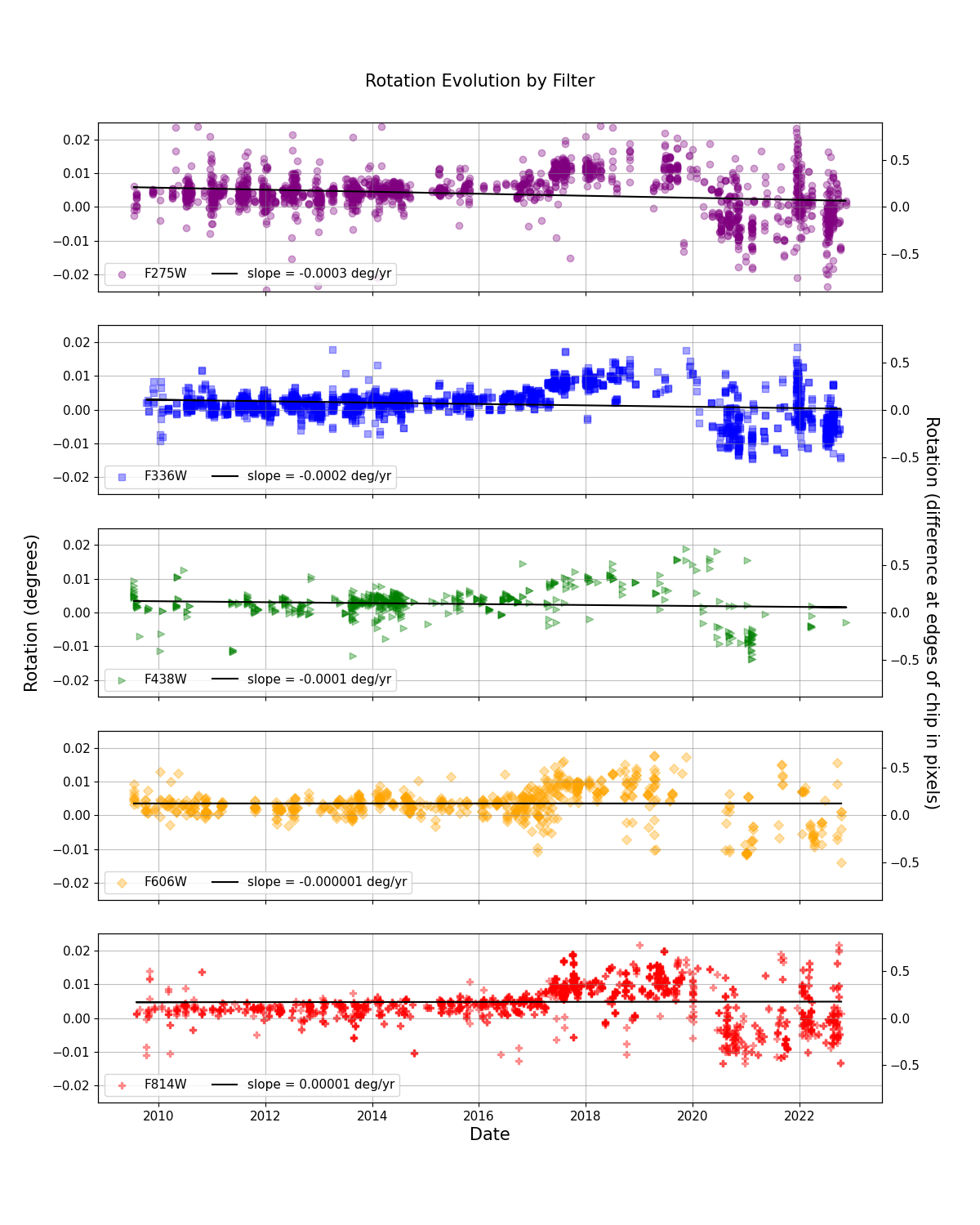}
    \caption{Rotation term offsets by filter: WFC3/UVIS to Gaia DR3. Each color and shape combination represents a separate filter. A line of best fit is plotted in black.}
    \label{fig:orient_subplot}
\end{figure}

%% file: plots/scale_plots.tex
\begin{figure}[!t]
    \centering
    \includegraphics[width=18cm]{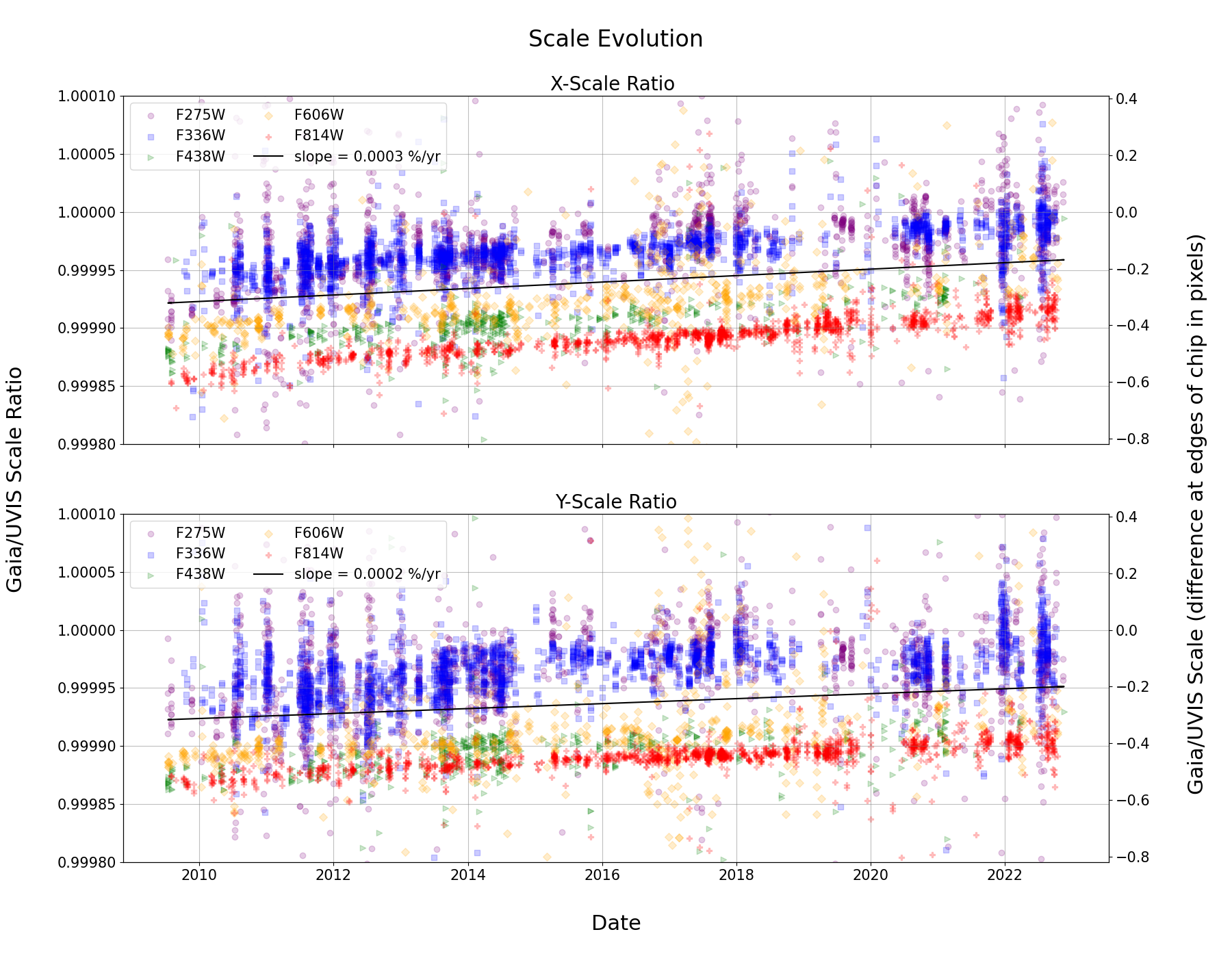}\caption    {Scale ratio: WFC3/UVIS to Gaia DR3. Each color and shape combination represents a separate filter. A line of best fit is plotted in black.}
    \label{fig:scale_overplot}
\end{figure}
\begin{figure}[!h]
    \centering
    \includegraphics[width=18cm]{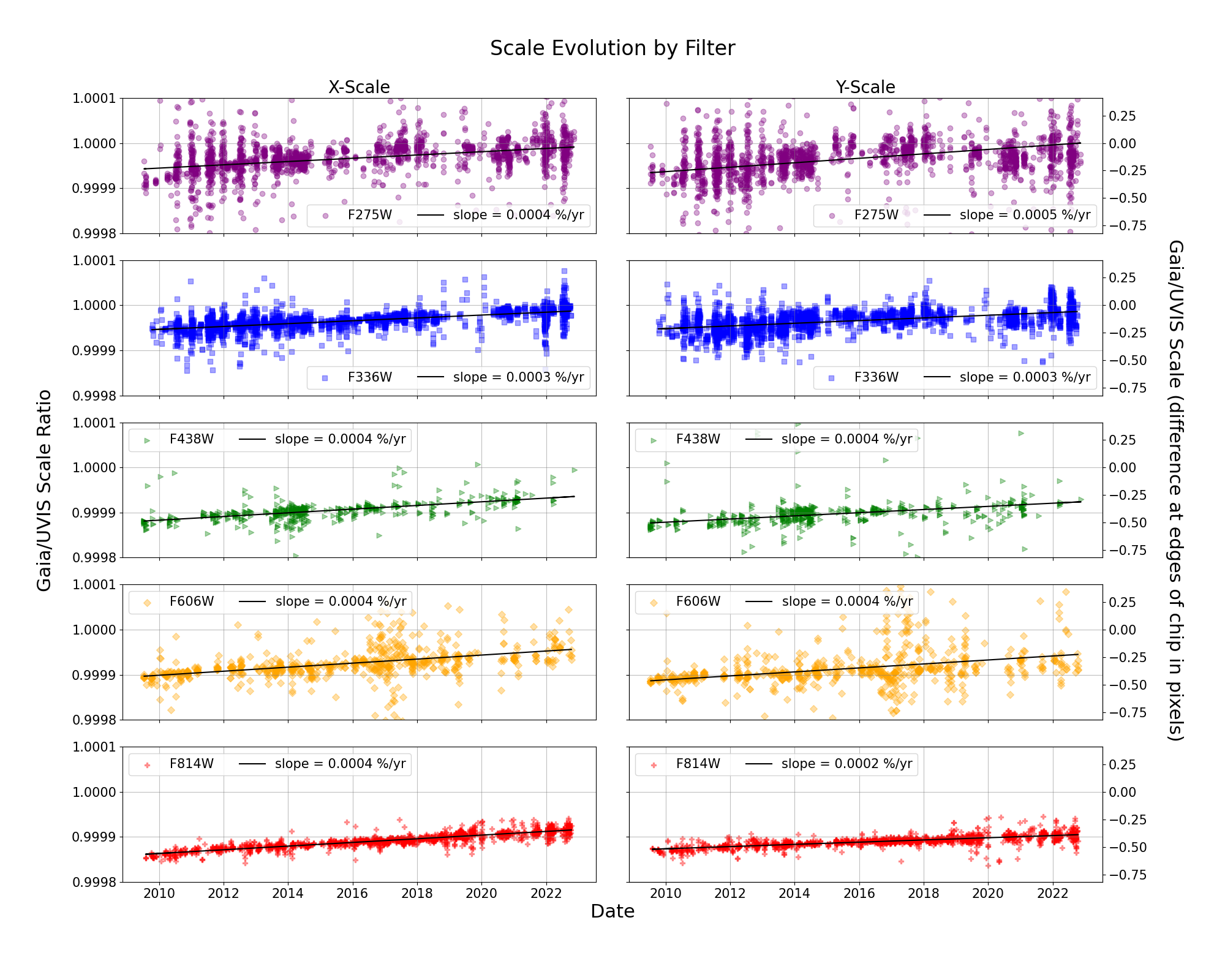}\caption    {Scale ratio by filter: WFC3/UVIS to Gaia DR3. Each color and shape combination represents a separate filter. A line of best fit is plotted in black.}
    \label{fig:scale_subplot}
\end{figure}

%% file: tables/scale_tables.tex
\begin{minipage}[!ht]{1\linewidth} \captionsetup{width=.9\linewidth}  \vspace{2ex}
\normalsize
    \centering 
    \def\arraystretch{1.3} 
    \begin{tabular}{c|c|c|c|c|} 
    \cline{2-5} 
      & \multicolumn{4}{c|}{\textbf{$\mathbf{X}$-scale}} \\ 
    \hline
\multicolumn{1}{|c|}{\textbf{Filter}} &  \specialcell{\textbf{Average Change} \\ (px)}  &  \specialcell{\textbf{RMS} \\ (px) } &  \specialcell{\textbf{Slope} \\ ($\mu$as px$^{-1}$yr$^{-1}$) } &  \specialcell{\textbf{ Mean Offset} \\ (pix) }  \\ \hline
         \multicolumn{1}{|c|}{All}&    0.15&   +/- 0.26&0.11 &-0.25\\ \hline 
         \multicolumn{1}{|c|}{F275W}&   0.2&   +/- 0.5&0.14 &-0.13\\ \hline 
         \multicolumn{1}{|c|}{F336W}&   0.16&   +/- 0.07&0.13 &-0.14\\ \hline 
         \multicolumn{1}{|c|}{F438W}&   0.21&   +/- 0.07&0.16 &-0.39\\ \hline 
         \multicolumn{1}{|c|}{F606W}&    0.2&   +/- 0.1&0.18 &-0.31\\ \hline 
         \multicolumn{1}{|c|}{F814W}&   0.20&   +/- 0.04&0.16 &-0.44\\ \hline

    \end{tabular}
    
\captionsetup{type=table}
\captionof{table}{X-scale, WFC3/UVIS to Gaia DR3. The maximum difference in pixels at edge of detector from 2009 to 2022, the RMS of this difference in pixels, and the slope of the line of best fit in $\mu$as px$^{-1}$yr$^{-1}$ are reported. We also report the mean scale offset, in  pixels.}
    \label{tab:x-scale}
\vspace{2ex}
\end{minipage}

\begin{minipage}[!ht]{1\linewidth} \captionsetup{width=.9\linewidth}  \vspace{2ex}
\normalsize
    \centering 
    \def\arraystretch{1.3} 
    \begin{tabular}{c|c|c|c|c|} 
    \cline{2-5} 
      & \multicolumn{4}{c|}{\textbf{$\mathbf{Y}$-scale}} \\ 
    \hline
\multicolumn{1}{|c|}{\textbf{Filter}} &  \specialcell{\textbf{Average Change} \\ (px)}  &  \specialcell{\textbf{RMS} \\ (px) } &  \specialcell{\textbf{Slope} \\ ($\mu$as px$^{-1}$yr$^{-1}$) } &  \specialcell{\textbf{ Mean Offset} \\ (pix) }  \\ \hline
         \multicolumn{1}{|c|}{All}&   0.11&   +/- 0.27&0.09 &-0.25\\ \hline 
         \multicolumn{1}{|c|}{F275W}&  0.3&   +/- 0.6&0.20 &-0.13\\ \hline 
         \multicolumn{1}{|c|}{F336W}&   0.15&   +/- 0.09&0.19 &-0.14\\ \hline 
         \multicolumn{1}{|c|}{F438W}&   0.2&   +/- 0.1&0.14 &-0.42\\ \hline  
         \multicolumn{1}{|c|}{F606W}&    0.2&   +/- 0.2&0.18 &-0.35\\ \hline 
         \multicolumn{1}{|c|}{F814W}&   0.12&   +/- 0.03&0.10 &-0.44\\ \hline

    \end{tabular}
    
\captionsetup{type=table}
\captionof{table}{Y-scale, WFC3/UVIS to Gaia DR3. The maximum difference in pixels at edge of detector from 2009 to 2022, the RMS of this difference in pixels, and the slope of the line of best fit in $\mu$as px$^{-1}$yr$^{-1}$ are reported. We also report the mean scale offset, in  pixels.}
    \label{tab:y-scale}
\vspace{2ex}
\end{minipage}

%% file: plots/skew_plots.tex
\begin{figure}[!b]
    \centering
    \includegraphics[width=18cm]{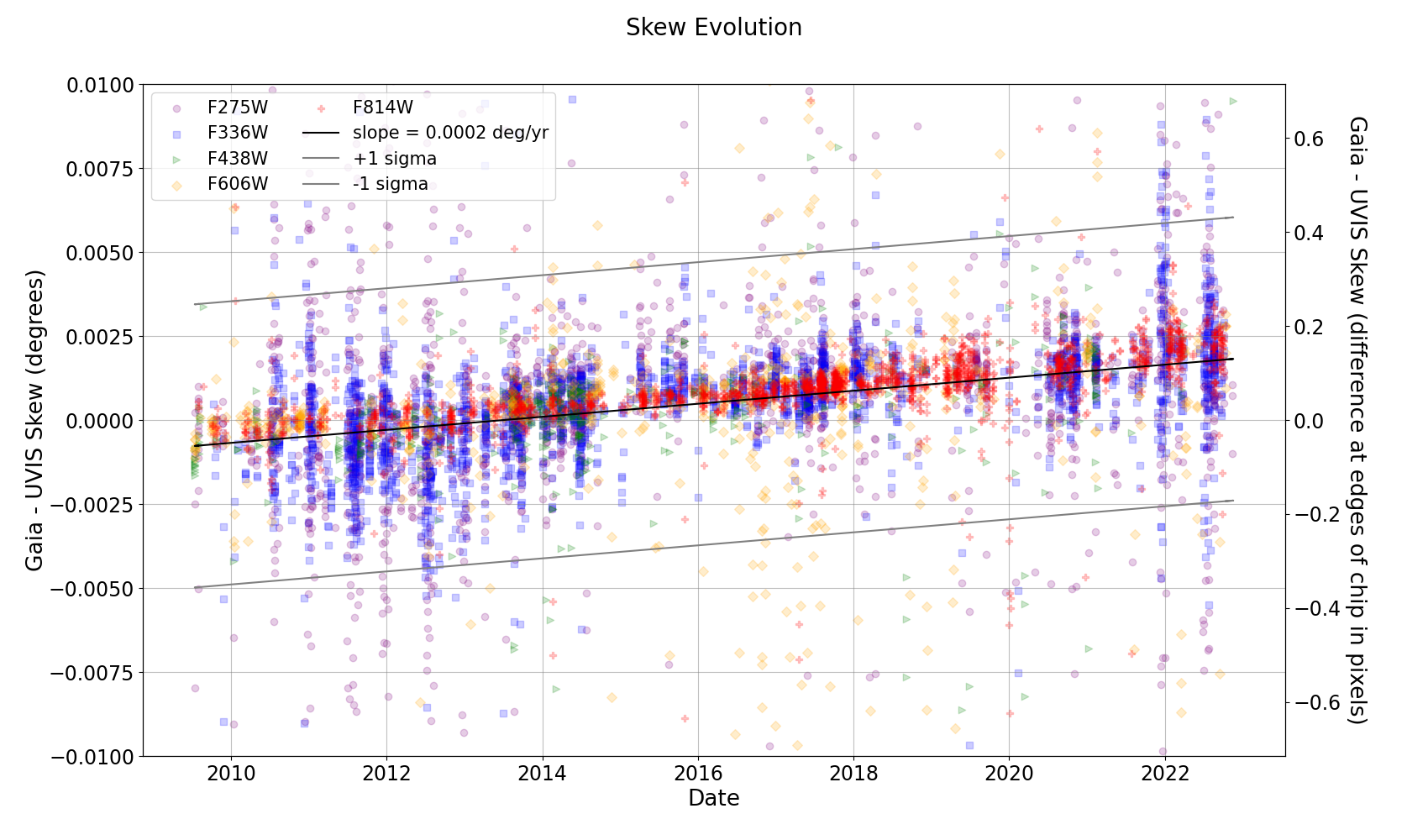}
    \caption{Skew term offsets: WFC3/UVIS to Gaia. Each color and shape combination represents a separate filter. A line of best fit is plotted in black.}
    \label{fig:skew_overplot}
\end{figure}

\begin{figure}[!h]
    \centering
\includegraphics[width=17cm ]{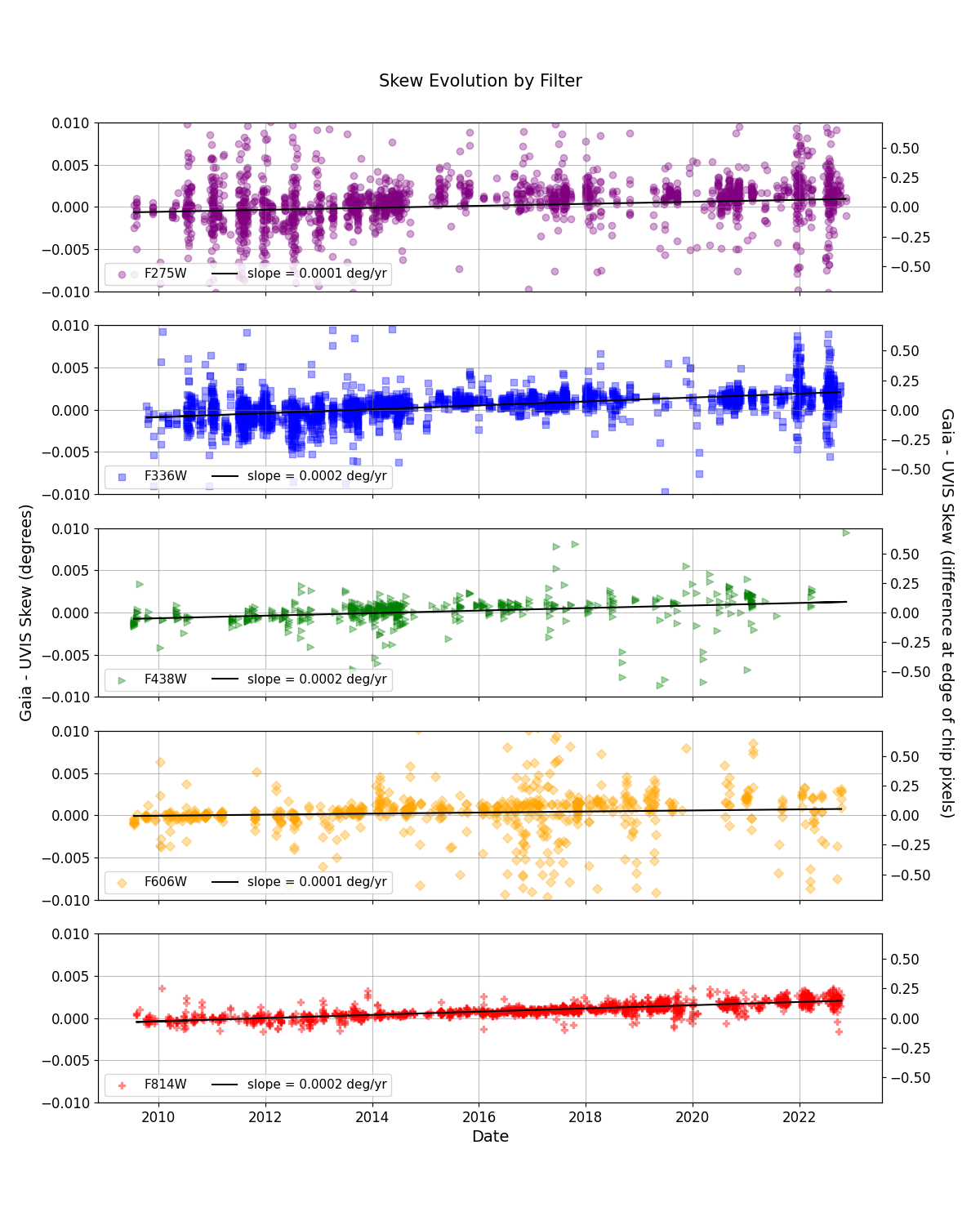}
    \caption{Skew term offsets by filter: WFC3/UVIS to Gaia DR3. Each color and shape combination represents a separate filter. A line of best fit is plotted in black.}
    \label{fig:skew_subplot}
\end{figure}

%% file: tables/skew_table.tex
\begin{minipage}[!ht]{0.9\linewidth} \captionsetup{width=.9\linewidth}  \vspace{2ex}
\normalsize
    \centering 
    \def\arraystretch{1.3} 
    \begin{tabular}{|c|c|c|c|} \hline
\textbf{Filter}  &  \specialcell{\textbf{Skew: Average Change} \\ (pix)}  &  \specialcell{\textbf{Skew: RMS} \\ (pix) } &  \specialcell{\textbf{Skew: Slope} \\ ($\frac{deg}{year})$ }  \\ \hline
         All& 0.2&+/- 0.3& 1.9E-4\\ \hline 
         F275W&  0.1&   +/- 0.9&1.2E-4\\ \hline 
         F336W&  0.2&   +/- 0.1&2.3E-4\\ \hline 
         F438W&   0.1&   +/-0.2&1.4E-4\\ \hline 
         F606W&   0.1&   +/- 0.2&6.2E-5\\ \hline 
         F814W&   0.2&   +/- 0.04&1.9E-4\\ \hline
    \end{tabular}
    
\captionsetup{type=table}
\captionof{table}{Skew, WFC3/UVIS to Gaia. The maximum difference in pixels at edge of detector from 2009 to 2022, the RMS of this difference in pixels, and the slope of the line of best fit in degrees/year are reported.}
    \label{tab:skew}
\vspace{2ex}
\end{minipage}